\documentclass[twocolumn, superscriptaddress]{revtex4-1}

\usepackage{graphicx}
\usepackage{amsmath}

\begin{document}

\title{Indirect measurement of three-photon correlation in nonclassical light sources}

\author{Byoung-moo Ann}
\affiliation{Department of Physics and Astronomy, Seoul National University, Seoul 151-747, Korea} 
\author{Younghoon Song}
\author{Junki Kim}
\author{Daeho Yang}
\author{Kyungwon An}
\email{kwan@phya.snu.ac.kr}
\affiliation{Department of Physics and Astronomy, Seoul National University, Seoul 151-747, Korea} 
\date{\today}

\begin{abstract}
We have observed the three-photon correlation in nonclassical light sources by using an indirect measurement scheme based on  the dead time effect of photon-counting detectors. We first developed a general theory which enables us to extract the three-photon correlation from the two-photon correlation of an arbitrary light source measured with detectors with finite dead times. We then
confirmed the validity of our measurement scheme in experiments done with the cavity-QED microlaser operating with a large intra-cavity mean photon number exhibiting both sub- and super-Poissonian photon statistics. The experimental results were in a good agreement with the theoretical expectation. Our measurement scheme provides an alternative approach 
for $N$-photon correlation measurement employing $(N-1)$ detectors and thus a reduced measurement time for a given signal-to-noise ratio, compared to the usual scheme requiring $N$ detectors.
\end{abstract}


\maketitle

\section{Introduction}\label{sec1}
Photon correlation measurement has played an important role in physical science. Since the second-order correlation (SOC) or two-photon correlation measurement was first devised by Hanbury-Brown and Twiss to resolve the size of stars \cite{Hanbury-Brown-57}, it has been utilized in various fields from ultra-cold atomic systems \cite{Folling-Nature05} to single molecule fluorescence \cite{Ververk-JCP03}. As more efficient measurement schemes have been developed \cite{Choi-RSI05}, many researchers has begun to address higher-order correlation functions in various aspects. M. Assmann {\em et al.} \cite{Assmann-Sci09} observed higher-order photon bunching effects in a semiconductor microcavity laser. 
Asymmetric temporal behavior of three-photon correlation in a strongly driven atom-cavity system was studied by M. Koch {\em et al.} \cite{Koch-prl11}. Higher-order photon correlation is also used in Doppler optical coherence tomography \cite{Huang-OL11} as well as in high-order ghost imaging \cite{Chan-OL09, Chan-OE10}. In matter wave experiments, the third-order correlation of Bose-Einstein condensates was investigated to confirm its Bose statistics \cite{Hodgman-Science11}.

One of the main benefits obtainable from the photon correlation measurement is the information on photon statistics. For instance, the normalized variance of photon statistics is simply described in terms of the SOC as $(\Delta n)^2 /\langle n\rangle-1=\left \langle n \right \rangle(g^{(2)}(0)-1)$. Here, $n$ is the photon number, $g^{(N)}(0)$ with $N=2$ is the second-order correlation and the left-hand side is usually called Mandel Q factor. The second-order- and furthermore the higher-order correlation functions give us useful information on photon statistics of a light source. In describing the statistical distribution of photon numbers, people frequently use the more generalized quantities called skewness and kurtosis, which are related with the third and forth-order correlations, respectively. 
These quantities give graphical and intuitive information on photon statistics, describing how much the distribution is asymmetrical and sharp, respectively. 

In photon correlation measurements, $N$ separate photodetectors are used for the $N$th-order correlation function in most configurations. For SOC measurement, Hanbury-Brown-Twiss-type measurement is often used. It may seem that this configuration would remove the detector dead time effect because two successive photons detected at separate detectors appear to be free from the dead time effect of each detector. Unfortunately, the dead time effect on SOC nevertheless exists even in this usual configuration and the distortion in SOC is known to depend on the higher-order ($N\ge 3$) correlation functions \cite{Schatzel-JOSAB89, Schztzel-APB86,Overbeck-rsi98,Flammer-AO97}. From this consideration, one can expect that it might be possible to extract the information on the higher-order correlations from the SOC measurement by analyzing the dead time effect. 
This possible new scheme to measure the correlation by using the detector dead time effect will be called `indirect measurement' throughout the paper.

However, most of the discussions in the literature \cite{Schatzel-JOSAB89, Schztzel-APB86,Overbeck-rsi98,Flammer-AO97} are restricted to the case where the light source is classical and the correlation time is much larger than the mean waiting time or the inverse of the photodetection flux. On the other hand, our previous study \cite{Ann-pra15} considered the dead time effect on non-classical light but neglected the higher-order correlations in deriving a correction formula. It is thus desirable to generalize all of these previous studies to cover both non-classical light sources and the higher-order correlations.

In this paper, we have theoretically derived an indirect measurement scheme for the third-order correlation by generalizing the previous studies on the dead time effect. We then applied it to the three-photon correlation measurement in the cavity-QED (quantum electro-dynamics) microlaser \cite{An-PRL94}. We have observed a relation between $g^{(2)}(0)$ and $g^{(3)}(0,0)$ experimentally, which well agrees with our theoretical expectation.

This paper is organized as follows. In Sec.\ \ref{sec2}, our indirect measurement scheme on the third-order correlation is theoretically discussed. We first show how the distortion of photodetection flux is connected with $g^{(2)}(0)$ and then extend the discussion to the distortion of $g^{(2)}(0)$ itself. This distortion eventually gives the information on the third-order correlation $g^{(3)}(0,0)$. In Sec.\ \ref{sec3}, we then prove the relation $3(1-g^{(2)}(0)) \simeq 1-g^{(3)}(0,0)$ in the cavity-QED microlaser when it operates with a large mean photon number. In the following section, we describe an experiment performed with the cavity-QED microlaser, where we extract $g^{(3)}(0,0)$ by using the indirect scheme and then show the observed $g^{(3)}(0,0)$ satisfies the above relation derived in Sec.\ \ref{sec3}. Concluding remarks will be given in Sec.\ \ref{sec4}.

\section{Theory of indirect measurement scheme}\label{sec2}

In this section, we derive how the photodetection flux and $g^{(2)}(0)$ are distorted under the detector dead time effect, which eventually provides a theoretical background for the indirect measurement scheme below. 

\subsection{$g^{(2)}(0)$ dependence of photodetection flux distortion}
   
\begin{figure}
\centering
\includegraphics[width=3.5in]{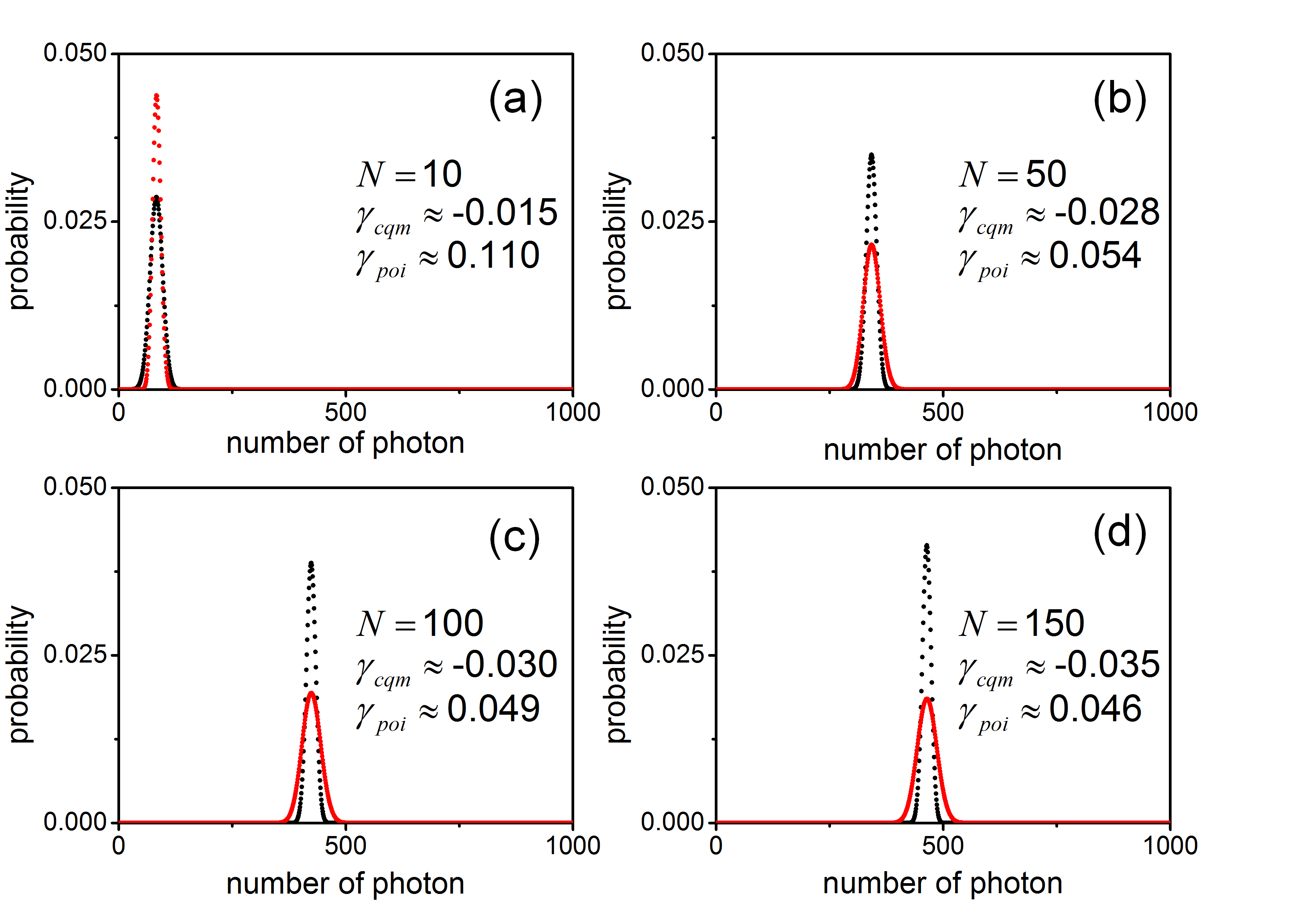}
\caption{(Color online) Illustration of photodetection events and dead time effect in a two-detector configuration. The black circles represent the observed photodetection events whereas the gray circles indicate the missed events caused by detector dead time, whose duration is marked by blue arrows. The correlation between the observed photodetection events (black circles in the first and third rows) corresponds to $g'^{(2)}(t)$. The second and forth rows are copies of the first and third rows, respectively, without distinction between the observed and missed photodetection events.}
\label{fig1}
\end{figure}

In this subsection, we first reveal that the dead time effect on photodetection flux is related with $g^{(2)}(0)$, which will be generalized to a relation between $g^{(3)}(0,0)$ and $g^{(2)}(0)$ in the subsequent subsection. 
Figure \ref{fig1} depicts photodetection events in a two-detector configuration where the black circles indicate observed photodetection events and the gray circles indicate missed events due to the detector dead time of duration $\tau$. The second and fourth rows refer to the corresponding dead-time-free photodetection events.
For further discussions, we hereby define 
\begin{equation}
g'^{(N)}(t_{12},...,t_{N-1,N}; \phi_{1},...,\phi_{N}; \tau_{1},..., \tau_{N}),
\label{eq3}
\end{equation}
an observed $N$-photon correlation function in the usual $N$-photodetector configuration when the dead-time-free photodetection flux and the dead time of $k$th detector are denoted as $\phi_{k}$ and $\tau_{k}$, respectively. Here, $t_{k-1,k}$ means the photodetection time delay between $(k-1)$th and $k$th detector. For example, the correlation between the black circles in the first and third rows in Fig.~\ref{fig1} is given by $g'^{(2)}(t;\phi_{st},\phi_{sp};\tau_{st},\tau_{sp} )$. When it comes to the correlation between the black circles in the first and second rows (for the identical incident photon streams), it can be expressed as  $g'^{(2)}(t;\phi_{st},\phi_{st};\tau_{st},0 )$.

The probability of finding dead-time-free photodetection events is apparently the sum of the probability to find the black circles and the probability to find the gray circles. The former is is related to the observed photodetection flux $\phi'(\tau)$ with a dead time $\tau$. 
The conditional probability to find a gray circle a time interval $t~(<\tau)$ after the appearance of a black circle is given by 
$g'^{(2)}(t; \phi, \phi; \tau, 0)$. Note that the correlation between the black and gray circles is the same as that between the observed events (black circles in the first and third rows) with the associated flux $\phi'(\tau)$ and the corresponding dead-time-free events (black circles in the second and forth rows) with the associated flux $\phi$. From this consideration, we derive the following relation,
\begin{equation}
\phi = \phi'({\tau}) +  \phi'(\tau) \cdot \phi \int_{0}^{\tau} g'^{(2)}(t; \phi, \phi; \tau, 0) dt.
\label{eq4}
\end{equation}
The first derivative of $\phi'({\tau})$ at $\tau=0$ is then given by
\begin{equation}
\left.\frac{d \phi'(\tau)}{d \tau} \right|_{\tau=0}  = -\phi^{2} g'^{(2)}(0;\phi,\phi;0,0) =  -\phi^{2} g^{(2)}(0),
\label{eq5}
\end{equation}
where we used the relation
\begin{equation}
\frac{d}{d \tau} {\int_{0}^{\tau} f(t,\tau)dt}  = f(\tau,\tau)+\int_{0}^{\tau} \frac{\partial f(t,\tau)}{\partial \tau}dt,
\label{eq5-1}
\end{equation}
in deriving Eq.~(\ref{eq5}) with $f(t,\tau)$ referring to an arbitrary multi-variable function of $t$ and $\tau$.
L. Mandel derived a similar formula to Eq.~(\ref{eq4}) for resonant fluorescence detection \cite{Mandel-prl83}. Only difference from our formula is that the integrand in Eq.~(\ref{eq4}) is $g^{(2)}(t)$, not $g'^{(2)}(t)$. This difference mainly comes from the fact that he neglected the effect of the higher order correlation.
   
\subsection{$g^{(3)}(0,0)$ dependence of SOC distortion}

In the following, we define the observed coincidence photodetection flux $\left \langle \phi_{st}'(\tau_{st}) \phi_{sp}'(\tau_{sp}) \right \rangle$, which refers to the number of simultaneous photodetection events at start and stop detectors per unit time when the start and stop detectors have dead time $\tau_{st}$ and $\tau_{sp}$, respectively. Then, $g'^{(2)}(0)$ measured in the two-detector configuration can be expressed as
\begin{equation}
g'^{(2)}(0;\phi_{st},\phi_{sp},\tau_{st},\tau_{sp})=\frac{\left \langle \phi_{st}'(\tau_{st}) \phi_{sp}'(\tau_{sp}) \right \rangle}{\phi'_{st}(\tau_{st})\phi'_{sp}(\tau_{sp})}.
\label{eq6} 
\end{equation}

\begin{figure}
\includegraphics[width=3.5in]{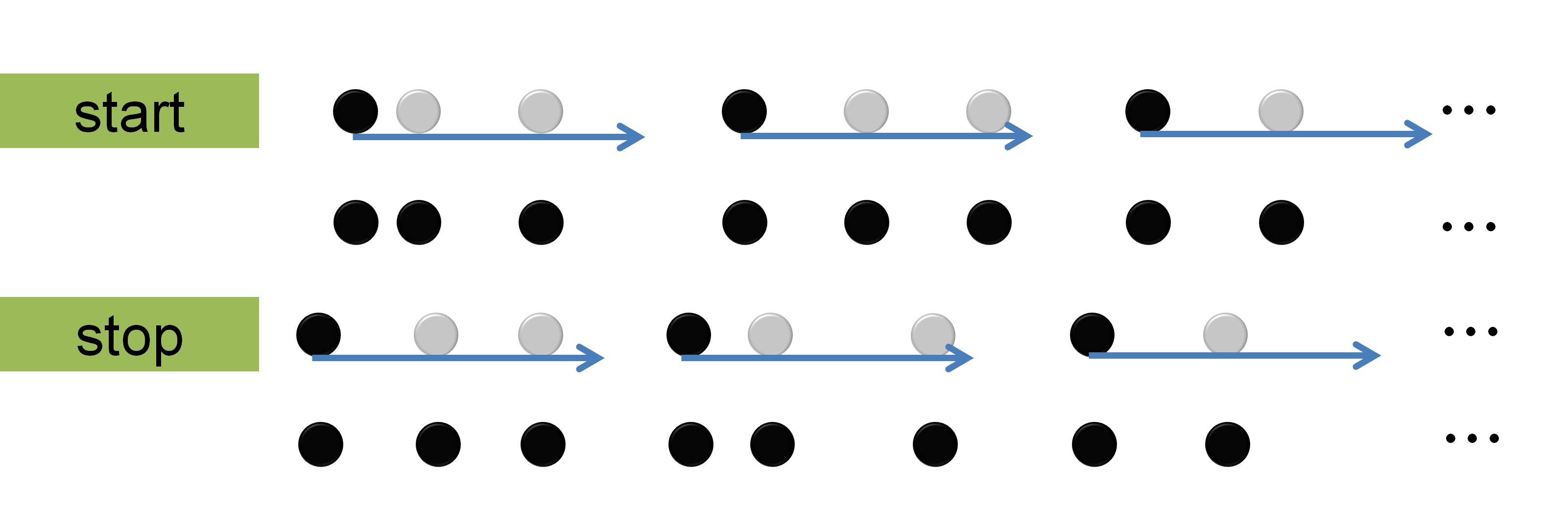}
\caption{(Color online) (a) Illustration of the missed coincidence photodetection when only stop detector has a dead time indicated by a blue arrow. For the photodetection scenario like a,b and c depicted above, we miss a coincidence event. (b) Third row, a duplicate of the second row, has been added to show that the correlation among a, b and c in (a) is the same as that in (b). Therefore, the correlation among these three events is given by $g'^{(3)}(t,0; \phi_{sp},\phi_{st},\phi_{sp};\tau_{sp},0,0)$.
}
\label{fig2}
\end{figure}

In Fig.~\ref{fig2}, we illustrate how the coincidence photodetection flux is distorted by the dead time while assuming $\tau_{st}=0$ for simplicity. The missing of coincidence events happens when the events a, b (both are detected events) and c (a missed event) satisfy the configuration depicted in Fig.~\ref{fig2}(a). The probability to have such three events is apparently same as the probability to have event a first and then events b and c simultaneously after time $t(<\tau_{sp})$ as depicted in Fig.~\ref{fig2}(b), which is given by the following expression
\begin{equation}
\phi_{st} \cdot \phi_{sp} \cdot \phi'_{sp}(\tau_{sp}) \cdot g'^{(3)}(t,0; \phi_{sp},\phi_{st},\phi_{sp};\tau_{sp},0,0),
\label{eq7}
\end{equation}
where $g'^{(3)}(t,0)$ approaches $g^{(3)}(t,0)$ as $\tau_{sp}$ decreases to zero. Accordingly, $\left \langle \phi_{st}'(0) \phi_{sp}'(\tau_{sp}) \right \rangle$ and the dead-time-free coincidence photodetection flux $\left \langle \phi_{st}\phi_{sp}\right \rangle=\left \langle \phi_{st}'(0)\phi_{sp}'(0)\right \rangle$ satisfy the following relation 
\begin{eqnarray}
\left \langle \phi_{st} \phi_{sp} \right \rangle&=&\left \langle \phi_{st}'(0) \phi_{sp}'(\tau_{sp}) \right \rangle + \phi_{st} \cdot \phi_{sp} \cdot \phi'_{sp}(\tau_{sp}) \nonumber\\
&  &\int_{0}^{\tau_{sp}} g'^{(3)}(t,0; \phi_{sp},\phi_{st},\phi_{sp},\tau_{sp},0,0) dt.
\label{eq7-1}
\end{eqnarray}
Note that Eq.~(\ref{eq7-1}) still holds after exchanging the indices start and stop. Differentiating both sides by $\tau_{sp} (\tau_{st})$ at $\tau_{sp} (\tau_{st})=0$ then gives
\begin{subequations}
\label{eq8}
\begin{eqnarray}
\left.\frac{d\left \langle \phi_{st}'(0) \phi_{sp}'(\tau_{sp})\right \rangle}{d\tau_{sp}} \right|_{\tau_{sp}=0} =-\phi_{st} \phi^{2}_{sp} g^{(3)}(0,0),\label{eq8a}\\
\left.\frac{d\left \langle \phi_{st}'(\tau_{st}) \phi_{sp}'(0) \right \rangle}{d\tau_{st}} \right|_{\tau_{st}=0} =-\phi_{sp} \phi^{2}_{st} g^{(3)}(0,0),\label{eq8b}
\end{eqnarray}
\end{subequations}
and thus, we can directly obtain the differentiation of Eq.~(\ref{eq6}), utilizing Eq.~(\ref{eq5}), as
\begin{subequations}
\label{eq9}
\begin{eqnarray}
\left.\frac{d g'^{(2)}(0;0,\tau_{sp})}{d \tau_{sp}} \right|_{\tau_{sp}=0}= \phi_{sp} \left([g^{(2)}(0)]^{2}-g^{(3)}(0,0) \right),\nonumber\\
\label{eq9a}\\
\left.\frac{d g'^{(2)}(0;\tau_{st},0)}{d \tau_{st}} \right|_{\tau_{st}=0}= \phi_{st} \left([g^{(2)}(0)]^{2}-g^{(3)}(0,0) \right).\nonumber\\
\label{eq9b}
\end{eqnarray}
\end{subequations}
In the linear approximation regime ($\phi\tau\ll 1$), we can express Eq.~(\ref{eq6}) as
\begin{eqnarray}
g'^{(2)}(0;\tau_{st},\tau_{sp})&-&g^{(2)}(0) \simeq 
 \left.\tau_{st}\frac{\partial g'^{(2)}(0;\tau_{st},\tau_{sp})}{\partial \tau_{st}}\right|_{\tau_{st,sp}=0}\nonumber\\
& & +\left.\tau_{sp}\frac{\partial g'^{(2)}(0;\tau_{st},\tau_{sp})}{\partial \tau_{sp}}\right|_{\tau_{st,sp}=0},
\label{linear}
\end{eqnarray}
and thus
\begin{equation}
g'^{(2)}(0)-g^{(2)}(0) \simeq \left\{[g^{(2)}(0)]^{2}-g^{(3)}(0,0)\right\} (\phi_{st}\tau_{st}+\phi_{sp}\tau_{sp}).
\label{eq9c}
\end{equation}

Combining Eq.~(\ref{eq4}) and Eq.~(\ref{eq9}) readily gives the second differentiation of $\phi'(\tau)$ and the result is
\begin{equation}
\left.\frac{d^{2} \phi'(\tau)}{d \tau^{2}} \right|_{\tau=0} = 2\phi^{3} g^{(3)}(0,0)+\phi^2 \left.\frac{dg^{(2)}(t)}{dt}\right|_{t=0},
\label{eq10}
\end{equation}
which clearly shows that the dead time effect on photodetection flux also depends on the higher-order correlation. 
The first order derivative of SOC function at zero time delay, $\left.\frac{dg^{(2)}(t)}{dt}\right|_{t=0}$, normally gives the inverse of a correlation time (an example is the cavity-QED microlaser), and thus Eq.~(\ref{eq10}) approaches the Flammer and Ricka's result \cite{Flammer-AO97} in the limit of a long correlation time ($\tau_w\ll\tau_c$), where $\tau_w$ is the mean waiting time in consecutive photodetections. 

Please note that $g'^{(2)}(0)$ is a function of $\phi_{st,sp}$ as well but we omit these in the above equations for simplicity of expression. We do not give further consideration for $g'^{(2)}(0)$ with a large $\tau$, for which the linear approximation does not work anymore. We should consider the forth- or even higher-order correlations in that case and it goes beyond the scope of the present work. 
We also neglect the dead time effect on the temporal dependence of SOC. It is because 
in the linear approximation regime the correlation time is much larger than the dead time and consequently the dead time effect on the correlation time is negligible.

We have not assumed anything about the photon source until now. 
Therefore, all the results here can be applied to any kinds of photon sources. Particularly, Eq.~(\ref{eq9c}), valid as long as $\phi\tau\ll 1$, would be broadly applicable as the information on $g^{(3)}(0,0)$ can be extracted from the first order coefficient of $g'^{(2)}(0)$ as a function of the detector dead time.

The discussion heretofore can be generalized to the case of $N$-detector configuration with only $N$th detector having the dead time $\tau_N$. Then, the following relation should hold:
\begin{eqnarray}
\left \langle \phi_{1}\ldots\phi_{N} \right \rangle&=&\left \langle \phi_{1}'(0) \ldots \phi_{N}'(\tau_{N})\right \rangle + \phi_{1} \ldots \phi_{N} \phi'_{N}(\tau_{N})  \nonumber \\
&\times& \int_{0}^{\tau_{N}} g'^{(N+1)}(0,...,t,0; \phi_{1},\ldots,\phi_{N-2},\phi_{N},\nonumber\\
& &\;\;\;\;\;\;\; \phi_{N-1},\phi_{N};0,\ldots,0, \tau_{N},0,0)dt
\end{eqnarray}
Deriving both sides by $\tau_{N}$ gives 
\begin{eqnarray}
& &\left.\frac{d g'^{(N)}(0;0,...,\tau_{N})}{d \tau_{N}}\right|_{\tau_{N}=0}\nonumber\\
& &=\phi_{N} \left[g^{(2)}(0)\cdot g^{(N)}(0)-g^{(N+1)}(0) \right].
\label{eq9-extended}
\end{eqnarray}
The subscripts $N$ can be replaced with arbitrary induces $i$ ($1\leq i \leq  N$ ), and thus 
\begin{eqnarray}
& &\left.\frac{d g'^{(N)}({0};0,...,\tau_{i},...,0)}{d \tau_{i}}\right|_{\tau_{i}=0}\nonumber\\
& &=\phi_{i} \left(g^{(2)}(0)\cdot g^{(N)}(0)-g^{(N+1)}(0) \right).
\label{eq9-generalized}
\end{eqnarray}
Finally, under the linear approximation $\phi_i\tau_i\ll 1$ we obtain
\begin{eqnarray}
& &g'^{(N)}({0})-g^{(N)}({0})\nonumber\\
& &\simeq \left[g^{(2)}(0)\cdot g^{(N)}(0)-g^{(N+1)}(0) \right] \left(\sum_{i=1}^{N} \phi_{i}\tau_{i}\right).
\label{eq9c-generalized}
\end{eqnarray}
This is a generalized form of Eq.~(\ref{eq9c}). Note that we omit the detector dead times in the argument of the observed correlation function for simplicity.

According to Eq.~(\ref{eq9c-generalized}), the constant term and the first order coefficient of $g'^{(N)}(0)$ as a function of dead times $\{\tau_i\}$ provide the information on dead-time-free $g^{(N)}(0)$ and $g^{(N+1)}(0)$, respectively. 
The detector dead time is fixed in general.  However, we can emulate various detector dead times on the record of observed photodetection events by deliberately deleting events. By plotting $g'^{(N)}(0)$ as a function of these emulated dead times, we can then extract both dead-time-free $g^{(N)}(0)$ and $g^{(N+1)}(0)$. The detailed procedure will be presented in Sec.\ \ref{sec4}.

\section{Photon correlation in the cavity-QED microlaser }\label{sec3}

The cavity-QED microlaser is a microscopic laser consisting of a high-Q optical cavity and a beam of two-level atoms traversing the cavity mode. The two-level atoms pumped by a conventional laser serve as a gain medium. It is an optical analogy of the micromaser \cite{Walther-PRL90}.
In the cavity-QED microlaser,
lasing is possible even when only one atom on average \cite{An-PRL94}, or a true single atom \cite{Lee-NC14} is present in the cavity. 
The coherent interaction between atoms and the cavity mode is still maintained even when the mean atom number in the cavity is much larger than unity \cite{An-JPSJ03}, exhibiting the novel properties such as quantum jumps \cite{Feng-Yen-pra06}, sub-Poisson photon statistics \cite{Choi-PRL06}, atomic \v{S}olc filter \cite{Hong-OE09, Seo-pra10} and quantum frequency pulling \cite{Hong-pra12, Hong-PRL12}, distinguished from the conventional laser. 

Quantum micromaser theory (QMT) \cite{Filipowicz-pra86} and semi-classical rate equation analysis predict various interesting features of the cavity-QED microlaser. The oscillatory gain function as the photon number of the cavity-QED microlaser represents the coherent interaction between the atom and the cavity mode, mainly accounting for the non-Poissonian properties of its intra-cavity photon statistics. The experiment done by W. Choi {\em et al.} \cite{Choi-PRL06} successfully observed these properties. 

In this section, we show that $g^{(2)}(0)$ and $g^{(3)}(0,0)$ satisfy a simple relation when the cavity-QED microlaser is operating with a large mean photon number. Let us first note that $g^{(3)}(0,0)$ can be expressed as follows in general

\begin{equation}
g^{(3)}(0,0)=\frac{{\left \langle n^3 \right \rangle}-3{\left \langle n^2 \right \rangle}+2{\left \langle n \right \rangle}
}{{\left \langle n \right \rangle}^3}.
\label{eq0}
\end{equation}
Meanwhile, skewness $\gamma$ of photon statistics is defined as
\begin{equation}
\gamma=\frac{{\left \langle n^3 \right \rangle}-3{\left \langle n^2 \right \rangle}{\left \langle n \right \rangle}+2{\left \langle n \right \rangle}^3}{{\left \langle \Delta n \right \rangle}^3}.
\label{eq0-1}
\end{equation}
We can rewrite Eq.~(\ref{eq0}) in terms of the skewness as
\begin{equation}
g^{(3)}(0,0)=1+\frac{3Q}{\left \langle n \right \rangle}-\frac{(3Q+1)}{\left \langle n \right \rangle^{2}}+\frac{(Q+1)^{3/2}}{\left \langle n \right \rangle^{3/2}}\gamma.
\label{eq1}
\end{equation}
In the following, we only consider the case that the Mandel Q factor is not close to zero since the cavity-QED microlaser shows non-Poissonian photon statistics except for few special atom numbers. In the case that $\left \langle n \right \rangle$ is much larger than $|Q|$ ({\em i.e.} $|1-g^{(2)}(0)| \ll 1 $), the third term in Eq.~(\ref{eq1}) becomes negligibly small compared to the second term. The skewness contained in the last term reflects the degree of asymmetry of the distribution. It becomes zero for a perfectly symmetrical distribution whose median and mean are the same such as a Gaussian or a delta-function distribution. With a large mean photon number and with a nearly symmetric photon number distribution, we can also neglect the last term in Eq.~(\ref{eq1}) compared with the second term.

\begin{figure}
\includegraphics[width=3.5in]{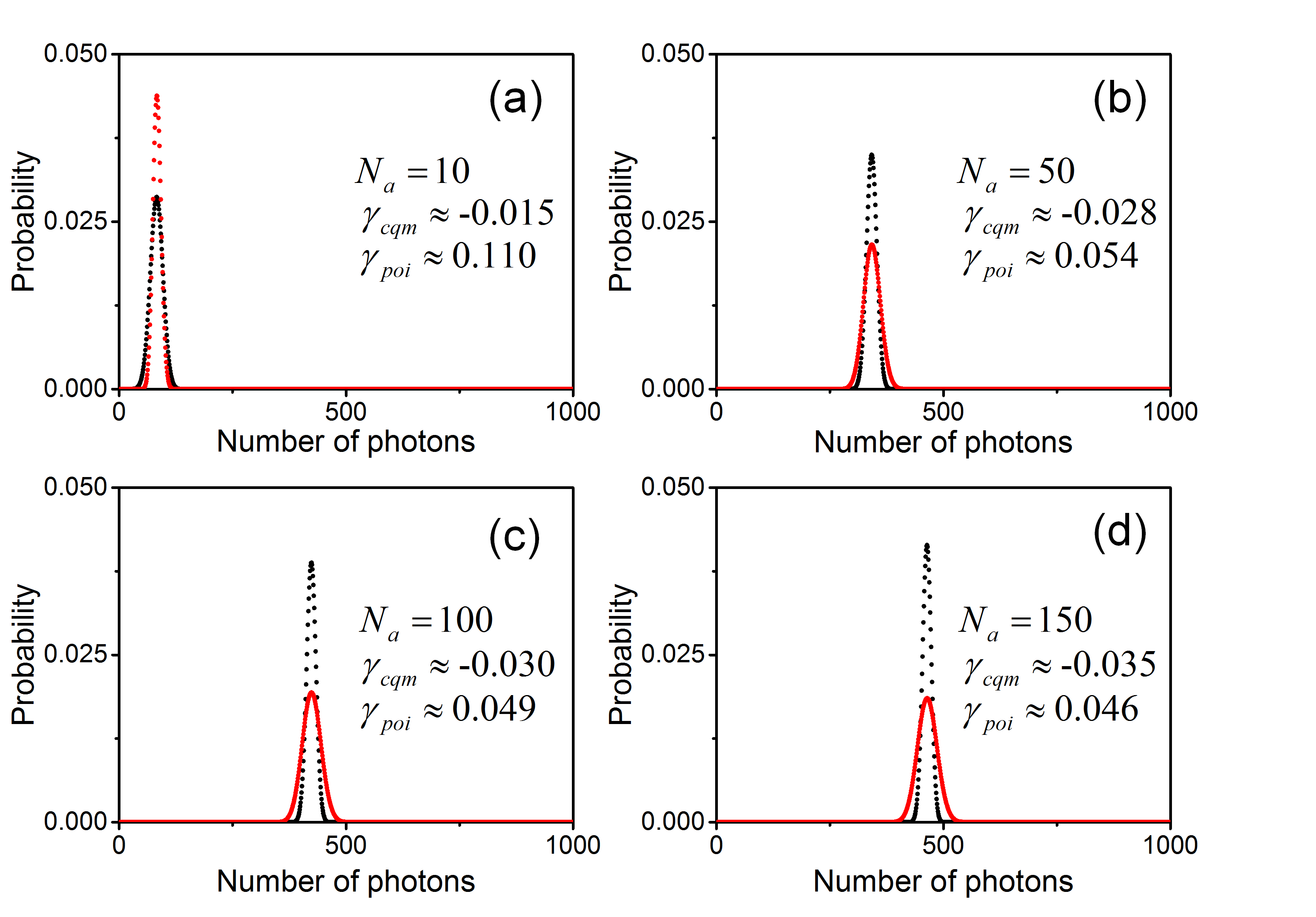}
\caption{(Color Online) Photon statistics of the cavity-QED microlaser operating in a large photon number region (black dots) well away from the regions where quantum jumps happen, compared to the Poissonian distributions (red dots) that have the same mean photon number. Here, $N_a$ refers to the mean atom number.
The photon statistics of the cavity-QED microlaser shows a near symmetric single-peak distribution both in super-Poissonian (a) and sub-Poissonian (b,c,d) regions when the mean photon number is sufficiently large. Skewness values of the cavity-QED microlaser are similar to or less than $\gamma_{poi}$ of the Poissonian distributions. The coupling constant and interaction time between the cavity mode and the impinging atoms assumed in the calculation were $2 \pi \times$190kHz and $0.1\mu$s, respectively, while the cavity linewidth was assumed by 138kHz. These operating parameters are basically the same as in Ref.\ \cite{Choi-PRL06}.}
\label{fig3}
\end{figure}

The photon statistics of the micromaser as well as the cavity-QED microlaser have been calculated in many previous studies \cite{Davidovich-RMP96, Scully-QO97}.
Under the condition that atom number is large enough to generate a large intra-cavity mean photon number but still insufficient to lead a quantum jump, all of the previous studies give a symmetric single-peak distribution, which implies the skewness may have a tiny value. Based on QMT, we calculated the skewness for the cavity-QED microlaser in the case when $\left \langle n \right \rangle$ is sufficiently large (Fig.~\ref{fig3}), 
where $\gamma_{cqm}$ is the skewness (black dots) of the cavity-QED microlaser photon statistics at various atom number  $N_a$ and $\gamma_{poi}= \left \langle n \right \rangle^{-1/2}$ is that of the corresponding Poisson distribution (red  dots) having the same mean photon number. 

Figure \ref{fig3} 
reveals that the order of the magnitude of $\gamma_{cqm}$ is similar to or less than that of $\gamma_{poi}$ when $\left \langle n \right \rangle$ is large enough. Therefore, we can now neglect the last term of Eq.~(\ref{eq1}) as well for the cavity QED microlaser as long as it operates well away from the regions where quantum jumps happen.
Equation (\ref{eq1}) is then approximated as
\begin{equation}
g^{(3)}(0,0)\simeq 1+\frac{3Q}{\left \langle n \right \rangle}
\label{eq2}
\end{equation}
for a large mean photon number. Equation (\ref{eq2}) then readily gives 
\begin{equation}
3(1-g^{(2)}(0)) \simeq 1-g^{(3)}(0,0)
\label{specific-relation0}
\end{equation}
because we have $g^{(2)}(0)=1+Q/\left \langle n \right \rangle$ by definition. The relation can also be expressed as 
\begin{equation}
[g^{(2)}(0)]^{3}\simeq g^{(3)}(0,0)
\label{specific-relation}
\end{equation}
since we already assume that $g^{(2)}(0)$ and $g^{(3)}(0,0)$ are close to unity. 
Direct calculation of $g^{(2)}(0)$ and $g^{(3)}(0,0)$ based on QMT also gives a consistent result with ours \cite{Hong-thesis}.
I
Meanwhile, QMT completely ignore the non-ideal effects including the cavity dissipation and atomic velocity distributions. For this reason, the validity of Eq.~(\ref{eq2}) might be questioned under the realistic conditions since the non-ideal effects would considerably change the photon statistics. 
However, we expect that there is no dramatical change at least in the symmetric shape of the photon number distribution based on the following reasoning.

Incoherent and inhomogeneous effects tend to reduce the oscillatory behavior of the gain function. One of the examples can be found in Ref.\ \cite{Choi-thesis}, where with inclusion of the atomic velocity distribution the amplitude of the gain function decreases as $\left \langle n \right \rangle$ increases. As these effects get severe, the gain function of the cavity-QED microlaser will approach that of the conventional laser whose photon statistics is Poissonian, and thus the skewness of the cavity-QED microlaser will further get close to $\gamma_{poi}$. Therefore, we can still have $\gamma_{cqm} \sim \left \langle n \right \rangle^{-1/2}$ and thus Eq.~(\ref{eq2}) under the non-ideal effects. To summarize, although the non-ideal effects evidently influence the photon statistics, Eq.\ (\ref{specific-relation0})
would still hold. 

By plugging Eq.\ (\ref{specific-relation}) into Eq.\ (\ref{eq9c}) we then obtain 
\begin{equation}
g'^{(2)}(0)-g^{(2)}(0)=\left\{ [g^{(2)}(0)]^{2}-[g^{(2)}(0)]^{3}\right\} (\phi_{st}+\phi_{sp})\tau,
\label{eq9d}
\end{equation}
where we assume $\tau=\tau_{st}=\tau_{sp}$.
Since $g^{(2)}(0)$ is closed to unity, Eq.~(\ref{eq9d}) is numerically equivalent to
\begin{equation}
 g'^{(2)}(0)\simeq g^{(2)}(0)\left \{1+\left[1-g^{(2)}(0)\right](\phi_{st}+\phi_{sp})\tau \right \}, 
 \end{equation}
 which is the result independently derived in Ref.\ \cite{Ann-pra15}.

The result obtained in this section is specific to the cavity-QED microlaser in the large photon number limit. Although the result is not general, it can still serve as a cross check reference for the indirect measurement method discussed in the previous section and the experimental results to be discussed in the following sections.

\section{Results and discussions}\label{sec4}

\subsection{$g^{(3)}(0,0)$ measurement of cavity-QED microlaser}

In typical photon-counting SOC measurements, the detector dead time is fixed for a given detector. 
Nonetheless,
we can simulate various detector dead times by computationally deleting photodetection events in the observed photodetection records. Even with this approach, we still miss the photodetection events near the origin within the physical detector dead time. However, if $g'^{(2)}(0)$ as a function of the dead time $\tau$ shows sufficiently linear variation near the origin, one can then approximate its variation in low order polynomials of $\tau$ and obtain reliable values of the constant term and the first order coefficient by performing a least-$\chi^{2}$ fitting. 

\begin{figure}
\includegraphics[width=3.5in]{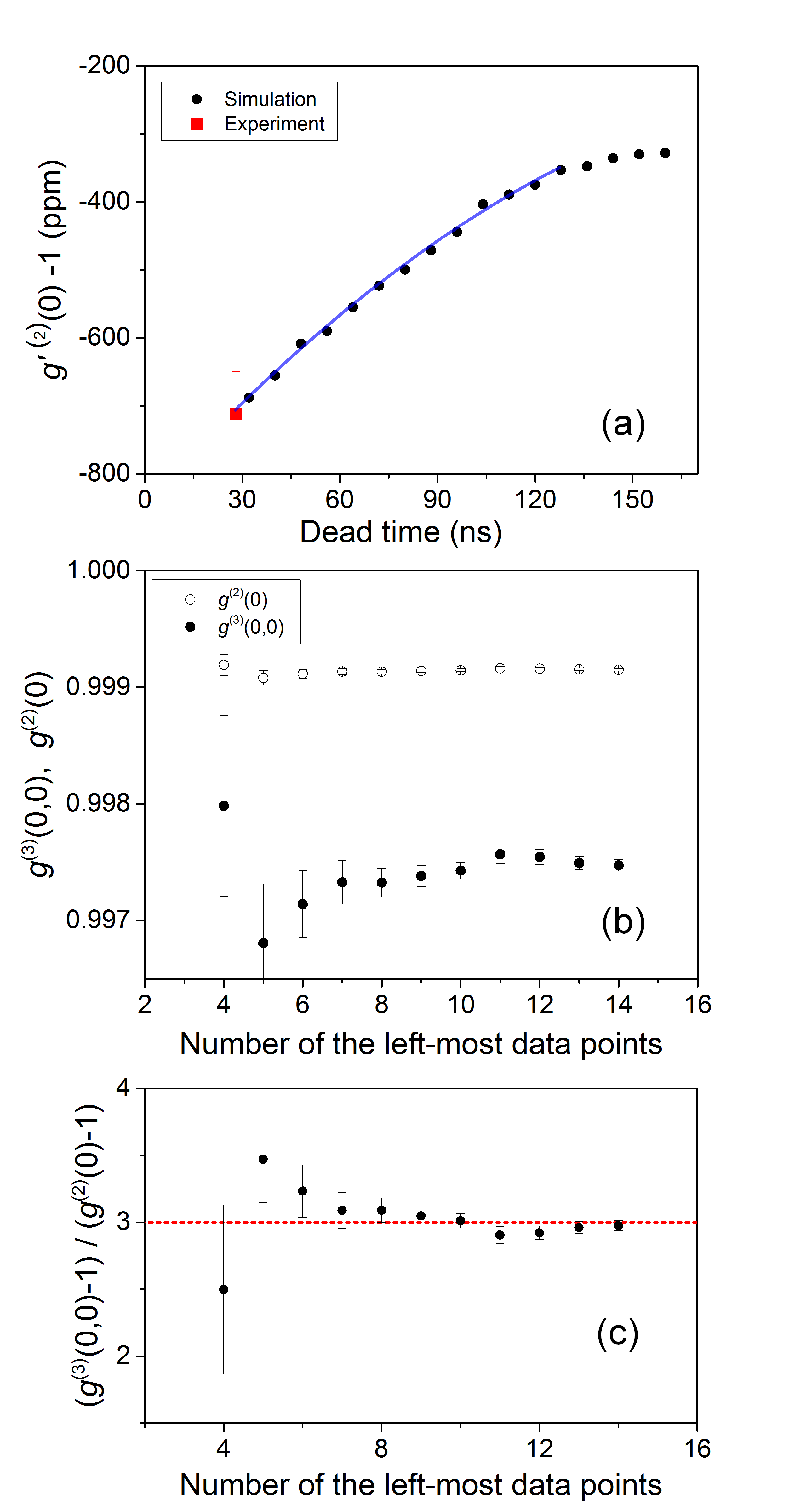}
\caption{(Color online) (a) Relation between $g'^{(2)}(0)$ and the dead time, observed by experiment and by prolonged dead times. A red square indicates experimentally observed result in the cavity-QED microlaser. Observed photodetection fluxes at each detector were 2.6 Mcps and 3.3 Mcps respectively. We simulated the prolonged dead time results (black circles) by deleting photodetection events from the actual data. The resulting $g'^{(2)}(0)$ is almost linearly increasing up to 128-ns dead time. Blue solid line indicates a least-chi-square fit curve when the data points up to the 14th participate in the fitting process. (b) Values of $g^{(3)}(0,0)$ (filled circles) and $g^{(2)}(0)$ (filled circles) obtained by performing 2nd-order polynomial fitting over the data points in (a) and by using Eq.~(\ref{eq9c}). Here, the horizontal coordinate indicates the number of data points participating in the fitting process. 
(c) Ratio of $g^{(3)}(0,0)-1$ to $g^{(2)}(0)-1$ based on the results in (b).
The more the number of the data points used in the fitting, the more reliable the results would become. Eventually the ratio converges to the theoretical expectation (dashed red line). Mean photon number in the cavity was roughly 600.
}
\label{fig4}
\end{figure}

Our experimental setup is basically the same as that in Ref.\ \cite{Ann-pra15}. Figure \ref{fig4}(a) shows $g'^{(2)}(0)$ as a function of the dead time measured in the regime of sub-Poisson photon statistics of the cavity-QED microlaser. A similar plot appeared in our previous work \cite{Ann-pra15}. 
The red square is obtained by the experiment when the mean photon number of the cavity-QED microlaser is approximately 600. The observed photodetection fluxes on start and stop photodetectors are 2.6 Mcps (mega counts per second) and 3.3 Mcps, respectively, and the dead time is 28 ns for both detectors. The dead-time-free photodetection flux was calculated by using Eq.~(2) in Ref.\ \cite{Ann-pra15} while approximately treating the output of the cavity-QED microlaser operating at a large mean photon number as coherent light.
We simulated the prolonged dead times, corresponding to the black circles, in the same way as in Ref.\ \cite{Ann-pra15}. The prolonged dead times imposed on both detectors were the same. At the photodetection fluxes given above, the emulated $g'^{(2)}(0)$ increases almost linearly until the 128-ns dead time. The fitting errors for the black circles are roughly 10\% of $|1-g'^{(2)}(0)|$.

In Fig.~\ref{fig4}(b)
we performed a least-$\chi^{2}$ fit of the data points in Fig.~\ref{fig4}(a) with a 2nd-order polynomial (a parabolic curve) from the red square to the 14th black circle corresponding to the 128-ns dead time. 
Here, $x$-axis refers to the number of the data points participating in the fit. 
Since the number of data points is limited in Fig.~\ref{fig4}(a), fitting with more than third-order polynomials tends to give a fitting error larger than the fitting parameters. 
From the values of $g^{(3)}(0,0)$ and $g^{(2)}(0)$ in Fig.~\ref{fig4}(b), we then obtain the ratio $[1-g^{(3)}(0,0)] / [1-g^{(2)}(0)]$ as shown in Fig.~\ref{fig4}(c). We observed that the relation $3[1-g^{(2)}(0)] \simeq 1-g^{(3)}(0,0)$ in Eq. (\ref{specific-relation0}) is well satisfied with a reasonably large number of the data points used in the fitting. The observed ratio with the fitting to the 14th data point was $2.98\pm0.07$. The error bars in Figs.~\ref{fig4}(b)-(c) indicate the fitting errors.

\begin{figure}
\includegraphics[width=3.5in]{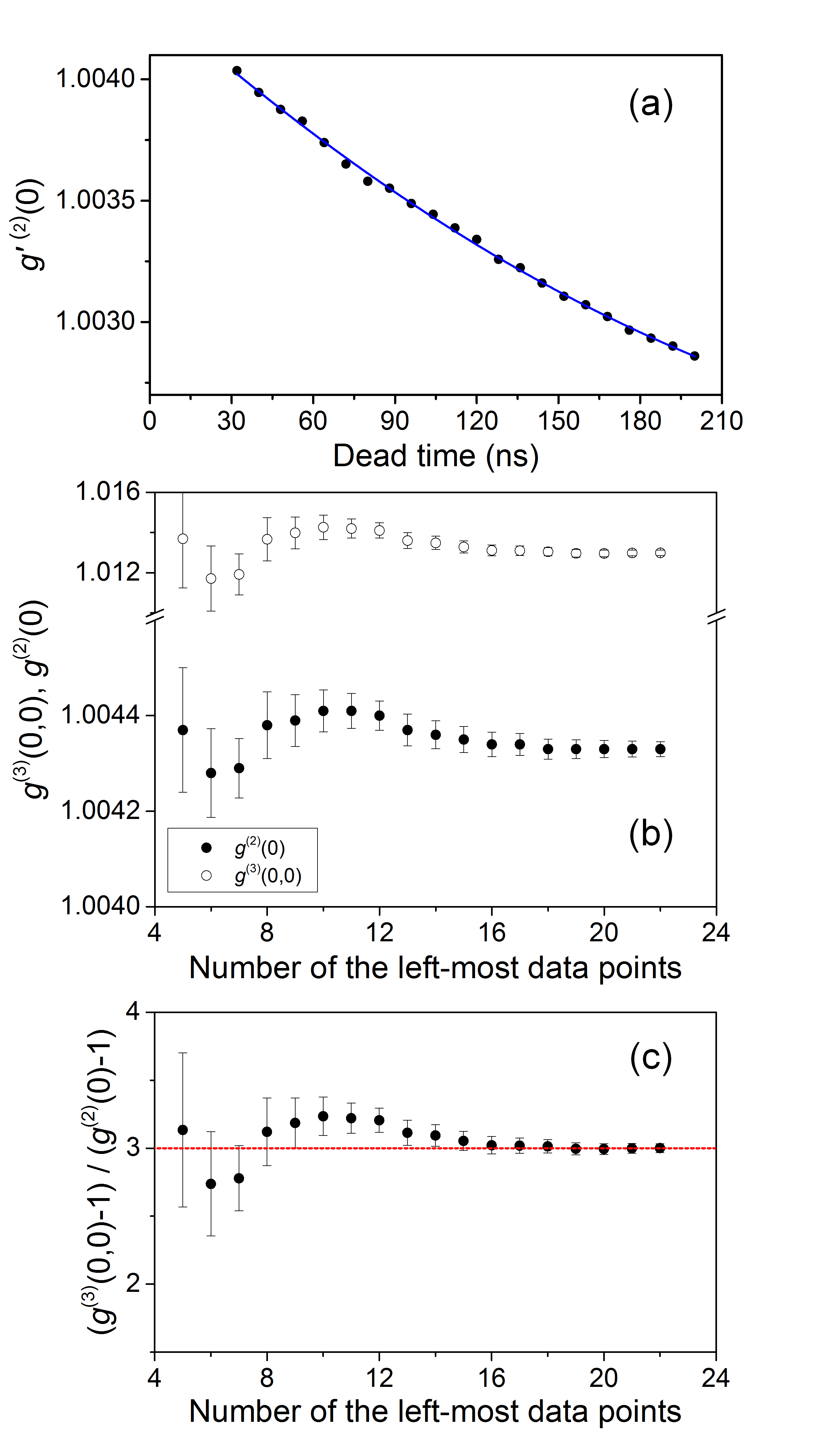}
\caption{(Color online) (a) Relation between $g'^{(2)}(0)$ and the dead time, observed in the super-Poissonian regime of the cavity-QED microlaser (black circles). 
(b) Values of $g^{(3)}(0,0)$ (open circles) and $g^{(2)}(0)$ (filled circles) obtained by performing 2nd-order polynomial fitting over the data points in Fig.~(\ref{fig5}a) and by using Eq.~(\ref{eq9c}).  Fitting errors for individual black circles are less than $1\%$ of $g'^{(2)}(0)-1$.  
Here, the horizontal coordinate indicates the number of data points participating in the fitting process. 
Blue solid line in (a) is the 2nd-order polynomial fitting curve when all the presented data points get involved in fitting process.
(c) Ratio of $g^{(3)}(0,0)-1$ to $g^{(2)}(0)-1$.
The obtained values of $|1-g^{(3)}(0,0)| / |1-g^{(2)}(0)|$ from the fitting with all data points is $3.00\pm0.03$. 
}
\label{fig5}
\end{figure}

We have also performed the $g^{(3)}(0,0)$ measurement with super-Poissonian light. Data in Fig.~\ref{fig5} correspond to $g'^{(2)}(0)$ measurements of the cavity-QED microlaser with a reduced mean intra-cavity atom number so as to generate super-Poissonian photon statistics. 
The observed photodetection fluxes were 870 kcps (kilo counts per second) and 1380 kcps for start and stop detector, respectively. 
The intra-cavity mean photon number for each data point was about 200.
We obtained $g^{(3)}(0,0)$ and $g^{(2)}(0)$ by the same way as in Fig.~\ref{fig4}. Fitting range was restricted up to the 200-ns dead time. The blue solid line is a 2nd-order polynomial fit curve. The value of $|1-g^{(3)}(0,0)| / |1-g^{(2)}(0)|$ obtained by the fitting is $3.00\pm0.03$ when the data points up to 200-ns dead time are involved in fitting, again well consistent with the theoretical expectation of Eq.~(\ref{specific-relation0}). 
The dark count was only 500cps and the contribution of afterpulse is only 0.5\% of the total photodetection flux, so we did not consider their effects in our data analysis.

\subsection{Advantage of the current indirect measurement scheme}
In the usual $N$-detector configuration for $N$th-order correlation measurement, signal-to-noise ratio (SNR) is given by  $\sqrt{(T_0/\tau_w)^N/(T_0/t_b)^{N-1}}=\sqrt{(T_0/\tau_w)(t_b/\tau_w)^{N-1}}$ \cite{Ann-pra15},
where $t_b$ is the bin time used in the calculation of the $N$-th order correlation function and $T_0$ is the total measurement time. 
The bin time $t_b$ should be small enough to resolve the temporal dependence of the correlation function, and thus it typically satisfies $t_b\ll \tau_w$. 
Therefore, in order to get a comparable SNR for the third-order correlation when compared with SOC measurement, the measurement time should be increased by a factor of $\tau_w/t_b$.
However, if the measurement time $T_0$ is limited by some technical reasons such as the finite oven life time as in the cavity-QED microlaser, the third-order correlation measurement will then be simply impossible. In this regard, our approach presented here provides a useful alternative to measure the higher-order ($N>2$) correlation. 

\section{Conclusion}\label{sec5}

We have developed a universally applicable theory which describes how the dead time distorts the second-order photon correlation in the usual two-detector configuration. Our formula relates $g^{(2)}(0)$ and $g^{(3)}(0,0)$ with $g'^{(2)}(0)$ as a function of the dead time, and thus, we can measure unknown $g^{(3)}(0,0)$ out of $g'^{(2)}(0)$. We call this new approach of obtaining $g^{(3)}(0,0)$ indirect measurement. In order to check the validity of our theory, we carried out an experiment with the cavity-QED microlaser in a large photon number regime and obtained $g^{(2)}(0)$ and $g^{(3)}(0,0)$ from $g'^{(2)}(0)$ by utilizing our formula.
Meanwhile, we have also shown that the cavity-QED microlaser satisfies $3(1-g^{(2)}(0)) \simeq 1-g^{(3)}(0,0)$ when it operates in large intra-cavity photon number. The observed $g^{(2)}(0)$ and $g^{(3)}(0,0)$ values in our experiments well agreed with the relation, thereby confirming the validity of our indirect measurement scheme.
Since our indirect approach is based on the second-order correlation measurement, it 
can greatly reduce the otherwise-much-longer measurement time for the third-order correlation for a desired signal-to-noise ratio, and thus it is particularly 
useful when the operation time of photon source is limited by some scientific and technical reasons.

\acknowledgements
This work was supported by a grant from Samsung Science and Technology Foundation under Project No.\ SSTF-BA1502-05.

\end{document}